Mixed cat states at low purity of light

N.I. Petrov

Scientific and Technological Centre of Unique Instrumentation of the Russian Academy of Sciences, 117342 Moscow, Russia

**ABSTRACT**. Pure states are usually used to observe quantum phenomena. In this study, we show that a quantum superposition of spatially displaced mixed cat states can be generated within an optical waveguide via nonparaxial unitary evolution of the initial low coherence (low purity) light beam. It is shown that highly mixed Schrodinger cat states can be observed at a well-defined propagation distance. The importance of the long-term decoherence and recoherence of the original wave packet in observing the mixed cat states is demonstrated. The Heisenberg and Schrodinger-Robertson uncertainty relations for mixed cat states are evaluated. We have demonstrated the feasibility of our method using accurate numerical simulations for the parameters of the source and optical waveguide available in practice.

The coherent superposition of different physical states [1] is a feature of quantum mechanics that distinguishes it most from classical mechanics. The coherent states (CS) of the harmonic oscillator have been used to demonstrate experimentally Schrodinger cat states in a cavity [2] and in a harmonic potential [3]. The coherent states represented by Gaussian wave packets were introduced for the first time by Glauber [4] for the one-dimensional stationary quantum oscillator in connection with quantum optics problems. Essentially, these states are analogous to the Gauss wave packets in a coordinate representation, which were constructed and studied by Schrodinger [5] for the quantum harmonic oscillator as part of his investigation of the relation between quantum and classical descriptions. The coherent states minimize the Heisenberg uncertainty relation, and they have the minimum possible width and diffraction angular divergence. The most used case has been the superposition of two distinguishable coherent states [6]. It was shown in [7] that Schrodinger's catlike states [5] are generated from an initial coherent state propagating through a Kerr nonlinear medium. The creation and measurement of coherent and Schrödinger-cat states have been performed in [8]. In [9] a method to produce quantum superposition of squeezed coherent states was proposed and experimentally demonstrated.

Despite the existence of the nonequivalence between classical and quantum mechanics and optics, very close analogies can be found between classical optics and quantum physics. Such analogies allow us to simulate quantum phenomena by means of light propagation in classical systems. It was shown in [10], that the coherent superposition of macroscopically distinguishable states can be generated via mode interference from an initial off-axis single wave packet. This opens up possibilities for visualizing quantum phenomena, such as the Schrodinger cat state, squeezing and collapse and revival, at a macroscopic level [10]. These phenomena are of great interest in many fields of physics such as quantum optics, quantum computation and precision measurements [11, 12].

It is generally assumed that pure states are necessary to observe quantum phenomena because mixed states destroy quantum effects. Creation of the superposition of two pure states requires cooling the system to near absolute zero to minimize noises. Therefore, the possibility of creating cat states from an initial thermal state is important [13-17]. Recently, it has been shown that quantum phenomena are possible even without extreme cooling [18]. It was experimentally shown that hot Schrodinger cat states can be generated within a microwave cavity at temperatures of up to 1.8 kelvin using interactions with a transmon qubit.

In this paper, we show that highly mixed Schrodinger cat states can be generated within an optical waveguide via nonparaxial unitary evolution of the initial low purity partially coherent beam at a well-defined distance. We have found that quantum interference can persist even in highly mixed states. This phenomenon is inherent in the general case of mixed states and, apparently, has not been studied in detail before. The importance of the relationship between the values of the coherence radius, the width of the incident partially coherent beam, and the waveguide gradient parameter in the observation of mixed cat states is emphasized.

## Theory

The Gaussian-Schell model (GSM) beams are familiar partially coherent sources which represent natural sources [19]. The coherence function of GSM beams is given by

$$\Gamma(x, x', 0) = I_0 exp\left\{-\frac{x^2+x'^2}{a_0^2} - \frac{(x-x')^2}{r_0^2}\right\}, \tag{1}$$

where $a_0$ is the beam width, $r_0$ is the coherence radius. Note that there is a close analogy between the coherence function and the quantum mechanical density matrix [20].

Here we consider the evolution of the coherence function (1) in the waveguide with the parabolic distribution of the refractive index in the transverse direction $x$:

$$n^2 = n_0^2 - \omega^2 x^2, \tag{2}$$

where $n_0$ is the refractive index on the axis, $\omega$ is the gradient parameter of the waveguide. Propagation of light in such a medium is similar to the behavior of the quantum-mechanical particle in a harmonic potential [10].

Mixed states representation of the coherence function can be used to study the behavior of a GSM beam in a graded-index waveguide. In this basis the coherence matrix takes diagonal form [21, 22]:

$$\Gamma(\tilde{x}, \tilde{x}', 0) = \sum_p \lambda_p \left|\Phi_p(\tilde{x})\right\rangle\left\langle\Phi_p(\tilde{x}')\right|, \tag{3}$$

where $\tilde{x} = x - x_0$, $\tilde{x}' = x' - x_0$, $x_0$ is the displacement of the incident beam relative to the waveguide axis, $\lambda_p = I_0\sqrt{\pi}\left(\frac{1}{r_0^2}\right)^p\left(\frac{1}{1/a_0^2+1/r_0^2+c}\right)^{p+1/2}$, and $\left|\Phi_p(\tilde{x})\right\rangle = \left(\frac{2c}{\pi}\right)^{1/4}\frac{1}{\sqrt{2^p p!}}exp(-c\tilde{x}^2)\mathcal{H}_p(\sqrt{2c}\tilde{x})$ are the Hermite-Gauss modes, where $c = \left(\frac{1}{a_0^4}+\frac{2}{a_0^2 r_0^2}\right)^{1/2}$.

Nonparaxial propagation of a light beam is described by the Helmholtz equation which is similar to the relativistic Klein-Gordon equation. The solution of the Helmholtz wave equation can be reduced to the solution of the Heisenberg equation for the operators, the average values of which determine the parameters of the investigated beam, e.g., the coordinate and width of the beam [10]. The nonparaxial evolution of the partially coherent beam in a graded-index waveguide (2) can be studied using the coherent states and coherent modes decomposition methods, as well as perturbation theory [22]. Here we apply the coherent mode decomposition method, which allows us to find precise solutions for analyzing the long-term nonparaxial evolution of wave packets. Evolution of coherence function can be represented as

$$\Gamma(\tilde{x}, \tilde{x}', z) = \sum_p \lambda_p \left|\Phi_p(\tilde{x}, z)\right\rangle\left\langle\Phi_p(\tilde{x}', z)\right|, \tag{4}$$

where the wave functions $\left|\Phi_p(\tilde{x}, z)\right\rangle$ are expanded into waveguide modal solutions

$$\left|\Phi_p(\tilde{x}, z)\right\rangle = \sum_m a_{pm}\left|\Psi_m(x)\right\rangle \exp\,(i\beta_m z), \tag{5}$$

where $\psi_m(x) = |m\rangle = \left(\frac{\pi}{k\omega}\right)^{-1/4} \frac{1}{\sqrt{2^m m!}} exp\left(-\frac{k\omega}{2}x^2\right) \mathcal{H}_m(\sqrt{k\omega}x)$ are the Hermite-Gauss modes, and $\beta_m = kn_0\sqrt{1 - 2\epsilon/n_0^2}$ are the propagation constants, where $\epsilon = \frac{\omega}{k}\left(m + \frac{1}{2}\right)$, $a_{pm}$ are the mode coupling coefficients given by the overlap integrals $a_{pm} = \langle m|p\rangle = T_p^m = \int \Psi_m^*(x)\, \Phi_p(x - x_0)dx$.

The overlap integrals can be calculated numerically or using recurrence relations [22]. Note that the mode spectrum is non-equidistant, which leads to effective nonlinearity. Considering only the first two terms in the expansion of the propagation constant $\beta_m(\epsilon)$, the model mathematically is equivalent to the evolution under Kerr medium Hamiltonian [10].

Substituting (5) into (4), we obtain

$$\Gamma(\tilde{x}, \tilde{x}', z) = \sum_p \lambda_p \sum_m \left|T_p^m\right|^2 |\psi_m\rangle\langle\psi_m| + \sum_p \lambda_p \sum_{m\neq n} T_p^m T_p^n |\psi_m\rangle\langle\psi_n| \exp[i(\beta_m - \beta_n)z]. \quad (6)$$

The first term in (6) describes a classical mixture that exhibits no interference properties, while the second term refers to the quantum coherence properties inherent in the cat state. At the coincident points $x = x'$, the expression (6) defines the intensity distribution $I(x, z)$.

**Results**

The concepts of entropy and quantum purity are closely related to the coherence properties of radiation. The entropy can be determined from the radiation coherence function [20, 22]. The entropy of the GSM beam can be determined as [20]:

$$S = -\sum_p \lambda_p \, ln\, \lambda_p, \quad (7)$$

where $\lambda_p = \frac{\langle p|\hat{\Gamma}|p\rangle}{Sp\hat{\Gamma}}$ is the probability of excitation of a given mode, i.e. the fraction of radiation energy carried by this mode.

The coefficients $\lambda_p$ depend on the coherence radius and the width of the incident beam, while the mode coupling coefficients $a_{pm}$ additionally depend on the parameters of the waveguide. Entropy reaches its minimum value in the case of coherent radiation and acquires a maximum value for incoherent radiation. Entropies as the measure of local coherence and global coherence (purity) can be considered.

The overall (global) degree of coherence, which is analogous to purity of a density matrix in quantum mechanics can be a useful quantity for characterizing partially coherent light [22]:

$$\mu^2 = \frac{Tr[\Gamma(x,x',z)\Gamma(x',x,z)]}{Tr\Gamma(x.x,z)Tr\Gamma(x',x',z)} = \frac{\sum_{m,n=0}^{\infty}\Gamma_{mn}\Gamma_{nm}^{+}}{\sum_{m=0}^{\infty}\Gamma_{mm}\sum_{n=0}^{\infty}\Gamma_{nn}}, \quad (8)$$

where $\Gamma_{mn} = \langle m|\hat{\Gamma}|n\rangle = \sum_p \lambda_p a_{pm} a_{pn}^*$.

In Fig. 1 the dependences of the overall degree of coherence (purity) and information entropy on the ratio of the radius of coherence to the width of the beam are shown.

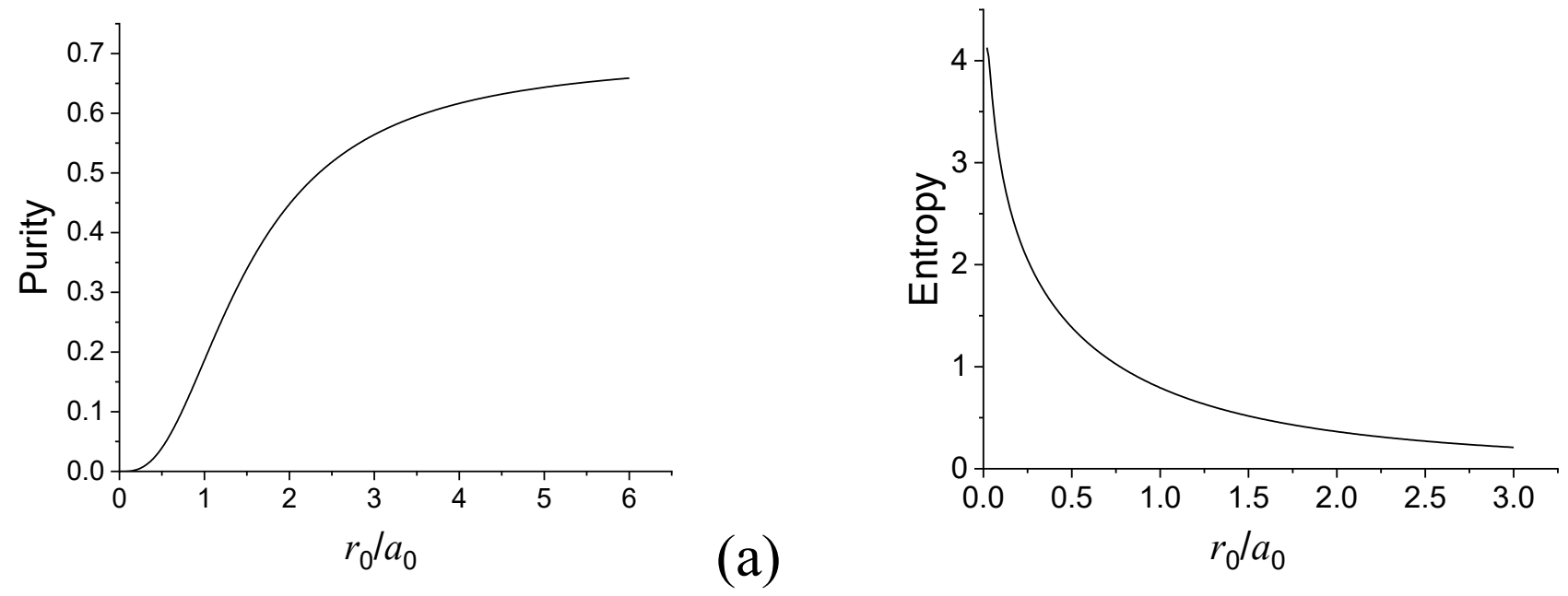

(a) (b)

FIG. 1. Purity (a) and entropy {b) as function of $r_0/a_0$.

The purity increases, and entropy decreases with the increasing coherence radius $r_0$. At a given value of $r_0$, purity decreases, and entropy increases with increasing beam width $a_0$. Note that the overall degree of coherence or purity, as opposed to the local degree of coherence, does not change during propagation along the graded-index waveguide.

The second-order moments of the geometric parameters of the GSM beam have significance to characterize the angular and spatial spread measures. In addition, they determine the radius of coherence and the degree of coherence of the beam. The coherence radius of the GSM beam (1) can be expressed as [22]

$$\frac{1}{r_c^2} = \frac{k^2(\sigma_x^2\sigma_p^2 - \sigma_{xp}^2 - 1/4k^2)}{2\sigma_x^2}, \quad (9)$$

where $\sigma_x$, $\sigma_p$, $\sigma_{xp}$ are the second-order moments of the geometric parameters, which define the beam width, angular spread and wavefront curvature radius, accordingly [22].

It is noteworthy that for $r_c \to \infty$, the right-hand side of (9), which defines the product of uncertainties, turns into zero, similarly for CS.

Figure 2 shows the change in the radius of coherence depending on the distance for the initial coherence radius $r_0 = 10$ μm.

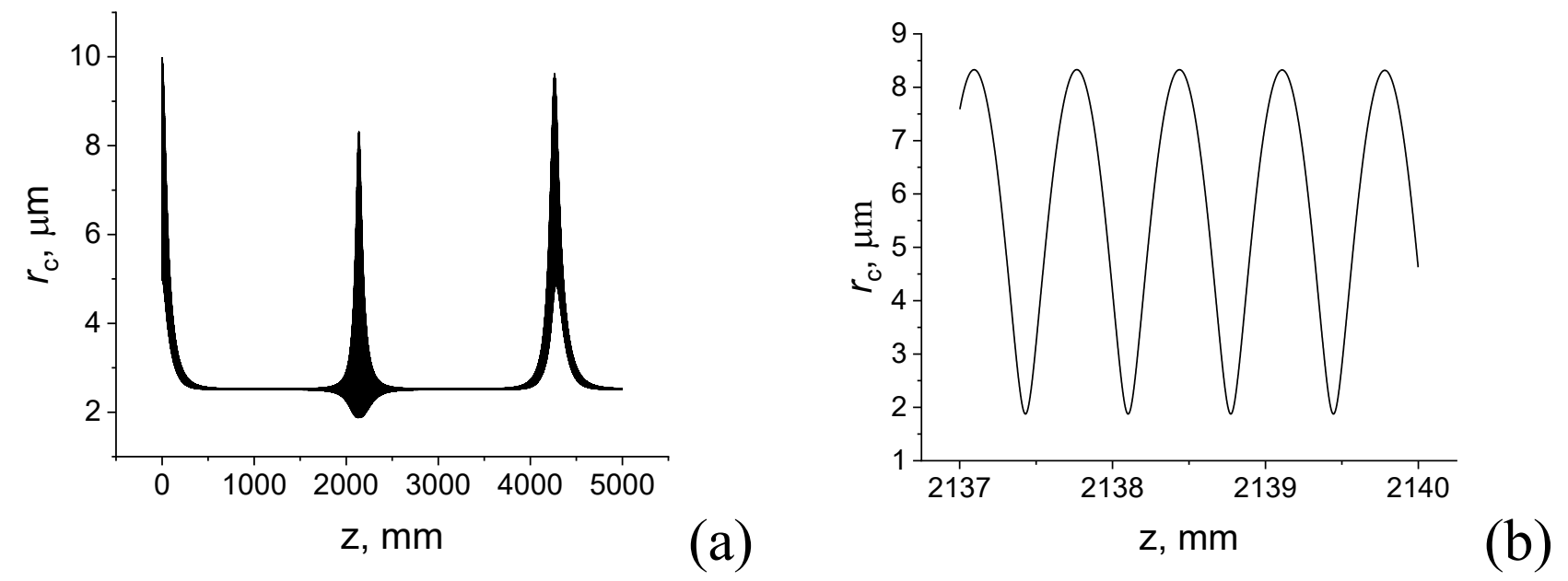

FIG. 2. (a) Coherence radius as function of distance. $a_0 = 10$ μm, $r_0 = 10$ μm, $x_0 = 10$ μm. (b) higher resolution plot

It can be seen that the radius of coherence decreases rapidly over short distance, but long-term recoherence occurs at a well-defined distance. Simulations show that the beam parameters undergo significant changes at nonparaxial evolution, including collapse and revival effects [22]. The striking feature of collapse and revival effects is the generation of squeezed states. In Fig. 3 the coefficient of the relative squeezing $\nu = \omega\sigma_x/\sigma_p$ as a function of a propagation distance is presented. A significant squeezing occurs at the midways between revivals (Fig. 3a). Note that the squeezing in classical optics has a clear physical meaning, and in the position (coordinate $x$) and momentum (the angle between the ray trajectory and the waveguide axis) space, it is defined as the ratio of the beam width and the angular divergence. Rapid oscillations of the squeezing occur in a small area, while squeezing is absent over most of the propagation distance. The enhancement of the squeezing is a signature of the appearance of the cat state. Superposition cat states are formed at distances where the squeezing coefficient reaches maximum values.

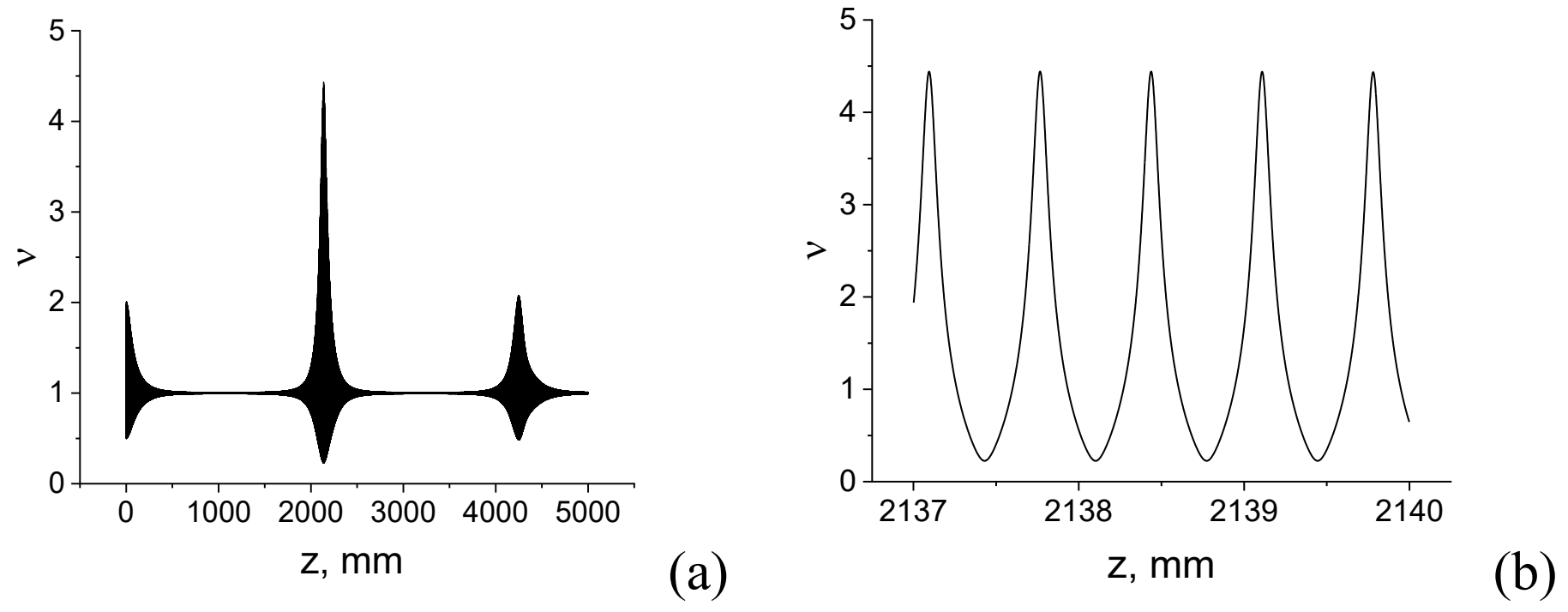

FIG. 3. Squeezing as function of distance. $n_0 = 1.5$, $\omega = 7 \cdot 10^{-3}$ μm$^{-1}$, $\lambda = 0.63\,\mu$m, $a_0 = 10\,\mu$m, $r_0 = 10\,\mu$m, $x_0 = 10$ μm. (b) Higher resolution plot showing oscillations with a period $L_{osc} = \pi n_0/\omega \approx 670\,\mu m$.

In Fig. 4, the evolution of uncertainty products are shown. These products acquire their initial minimum values at every revival growing to higher values in between revivals.

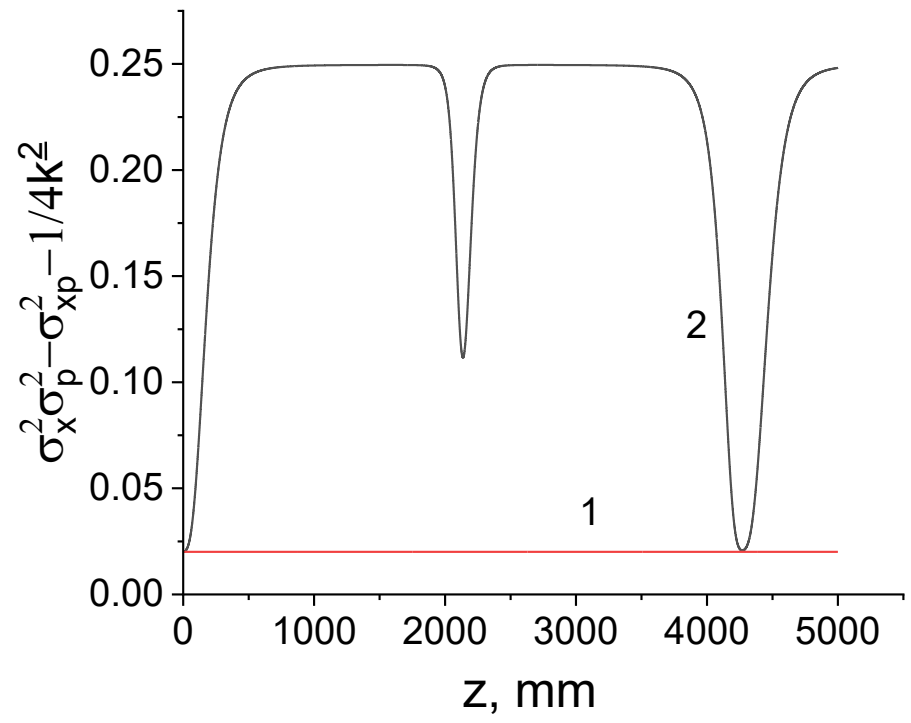
(a)

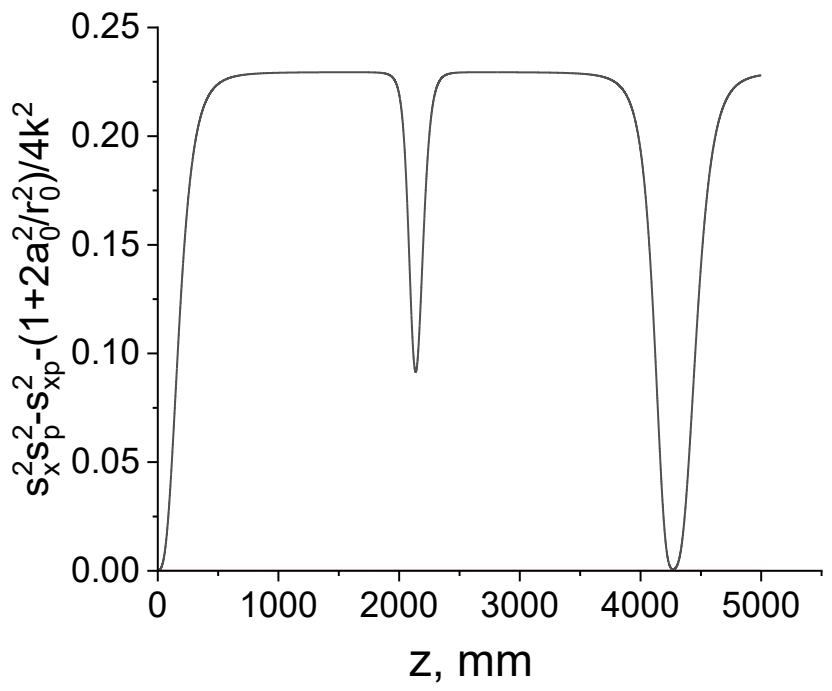
(b)

FIG. 4. Evolution of the uncertainty products of Schrodinger-Robertson (a) and Eq. (10) (b) for $x_0 = 10$ μm, $a_0 = 10$ μm, $r_0 = 5$ μm. 1 – paraxial, 2 – nonparaxial.

At the revivals, the uncertainty relations of Heisenberg, $\sigma_x^2\sigma_p^2 \geq \frac{1}{4k^2}$, and Schrodinger-Robertson [23], $\sigma_x^2\sigma_p^2 - \sigma_{xp}^2 \geq \frac{1}{4k^2}$ become an equality for the original coherent beam [10]. For the mixed states the generalized Schrodinger-Robertson uncertainty relation has the form [22]:

$$\sigma_x^2(z)\sigma_p^2(z) - \sigma_{xp}^2(z) \geq \frac{1}{4k^2}\left(1 + \frac{2a_0^2}{r_0^2}\right). \quad (10)$$

Note that for the mixed state (1) equality in (10) is valid, i.e. it is as close as possible to the mixed classical state. The uncertainty relation (10) assumes a minimum value at the revival distances, manifesting the classical features of the wave packet. While the initial mixed state minimizes the uncertainty relation (10), cat states do not minimize it, indicating that a cat state is a nonclassical object.

Nonlinear phase distortions caused by nonparaxiality destroy the initial field distribution. The most remarkable thing is that the disrupted field distribution transforms to a cat state at a well-defined distance. An off-axis wave packet of mixed states is transformed into a coherent superposition of two macroscopically distinguishable mixed states at the midways between revivals. Namely, the mixed cat states are formed at a distance $z_{cat} \cong z_{rev}/2$, where the revival interval $z_{rev} = \pi n_0/\eta$, where $\eta = \frac{\omega^2}{kn_0^2}$. In Fig. 5 the intensity profiles of the mixed cat states at different distances are shown.

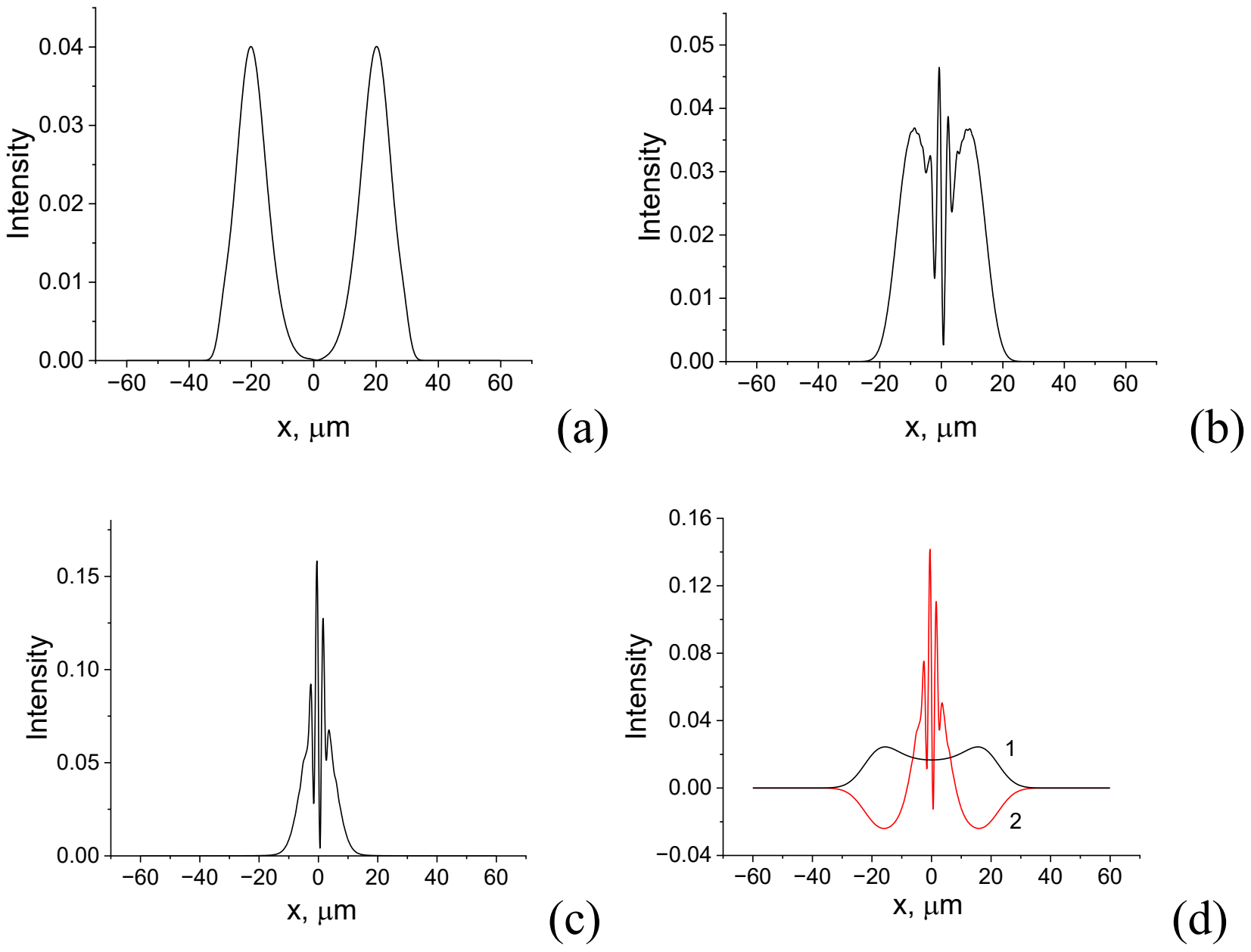

FIG. 5. Wave packet intensity distributions: $x_0 = 20$ μm, $a_0 = 10$ μm, $r_0 = 5$ μm, $\lambda = 0.63$ $\mu$m. (a) cat state at z =2115000 μm; (b) z =2115220 μm, (c, d) z =2115320 μm, 1 – first term from (6), 2 – second term from (6).

It can be seen that the intensity distribution consists of regular and oscillatory parts (Fig. 5d). While the regular part is formed by diagonal elements in (6), the oscillatory part refers to the cross elements that cause interference fringes (Figs. 5b, 5c). It is noteworthy that the cross terms in (6) do not contribute to the total power but only affect the intensity distribution profile.

Rapid collapse and revival of cat states with a period of $L_{osc} = \pi n_0/\omega$ occur resulting in interference fringes between the appearing of the superposition states. These short-term periodic oscillations show a coherent superposition rather than a simple mixture of two wave packets. Indeed, the incoherent superposition of the two initial mixed states (classical mixture) does not create interference fringes at $z = z_{cat}$ (Fig. 6d). Conversely, the mixed cat state is a coherent superposition of two low-coherence wave packets that interact to form interference fringes. Every individual initial wave packet generates cat states and interference patterns with similar mirror-symmetric profiles (see Figs. 6a and 6b). However, it turns out that they are slightly displaced in space (Fig. 6c). This leads to a smoothing of the overall intensity profile and to the complete disappearance of interference fringes (Fig. 6d). Thus, while the coherent cat state is generated from one initial mixed state, the statistical mixture is generated from two initial mixed states.

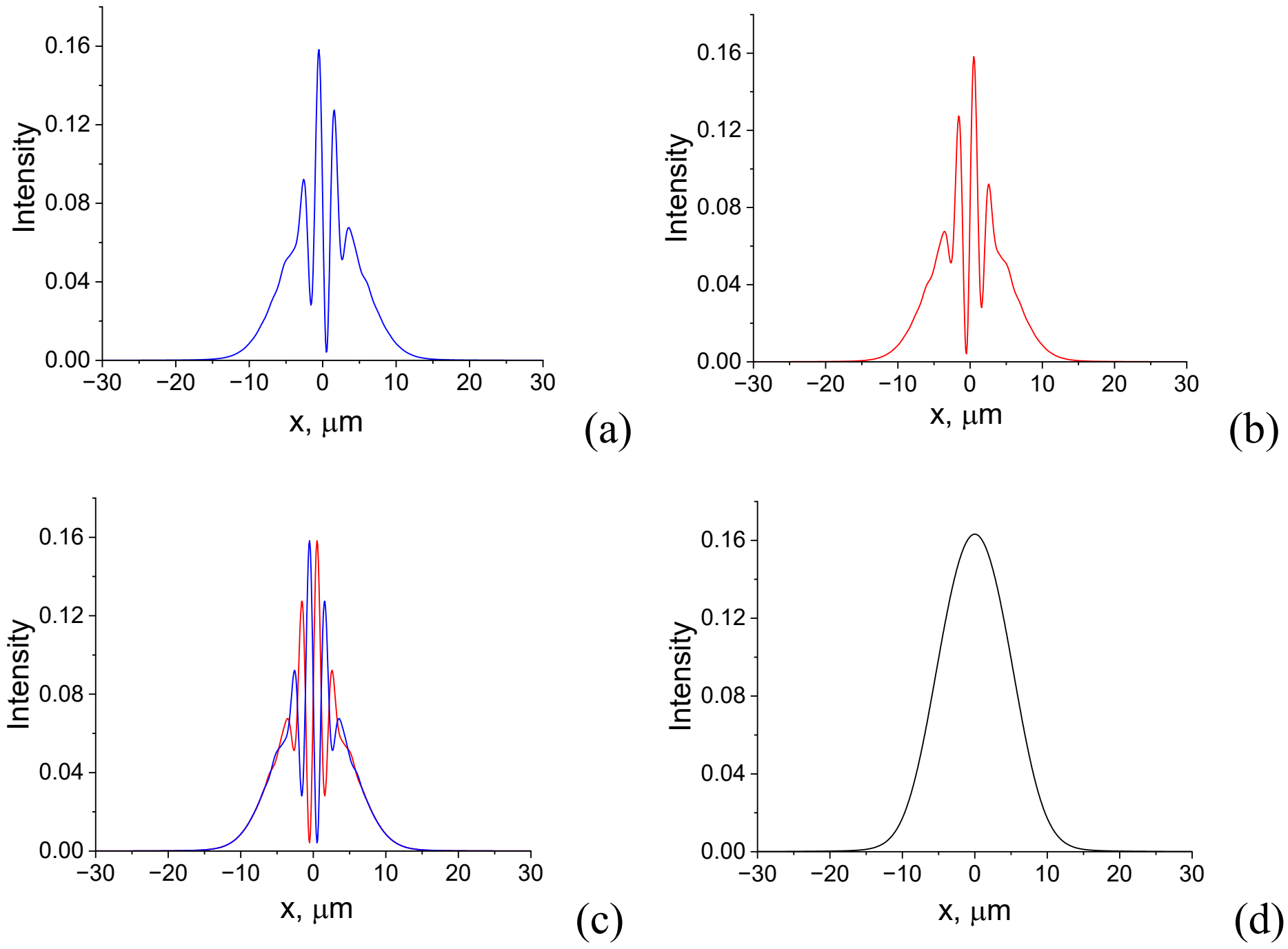

FIG. 6. (a, b) Intensity profiles of cat states created by initial mixed states with offsets of the opposite sign $x_0 = 20$ μm and $x_0 = -20$ μm, accordingly; (c) combined profiles; (d) overall intensity profile. z =2115320 $\mu$m, $a_0 = 10$ μm, $r_0$ = 5 μm, $\lambda = 0.63$ $\mu$m.

**Discussion**

Thus, the highly mixed Schrodinger cat states (a quantum superposition of macroscopically distinguishable states) can be generated by nonparaxial propagation of an initial partially coherent beam of low purity in an optical graded-index waveguide at a well-defined distance. The interference fringes exhibited by a mixed Schrodinger cat state are generated by the pure states from the statistical mixture. It is shown that mixed cat states can be observed with an initial purity of only 0.04 for an average number of waveguide excited modes of 12 and an off-axis displacement of 20 μm. For the observation of cat states, the width of the original beam $a_0$ should be close to the waist of the fundamental mode of the waveguide $w_0 = (2/k\omega)^{1/2}$. This information would be useful in experiments to investigate macroscopic quantum superposition. The importance of coherence properties in the analysis of a mixed cat states was also demonstrated in [13]. It was shown that the coherence length decreases with temperature and thus makes the detection of hot Schrodinger cat states more difficult.

The results obtained are well explained by numerical simulations. Note that the cat states formation and interference features occur only at a nonparaxial evolution and disappear at a paraxial propagation. Nonparaxial propagation of an optical beam is described by an equation similar to the relativistic Klein-Gordon equation. This indicates that the cat state is a feature of relativistic quantum mechanics.

Our precise calculations, based on an exact theory, show that the long-term recoherence of the wave packet is an obvious signature of the appearance of the cat state. Indeed, the degree of coherence and the radius of coherence increase sharply at a well-defined distance $z_{cat}$. This distance can be reduced by increasing the wavelength of the source and the gradient parameter of the waveguide.

Note that optical systems have an advantage over condensed matter systems due to the relative purity and simplicity of optical components (waveguides and partially coherent sources). Optical waveguides with sufficiently low dissipation are currently available. Unlike the special Yurke-Stoler Hamiltonian [7], which is difficult to realize in practice, effective nonlinearity is created here due to the non-equidistant spectrum of modes in a graded-index waveguide.

Future research may be related to the consideration of mixed cat states in an optical fiber when the original wave packet is partially coherent and partially polarized. This will allow us to analyze the effects caused by the spin-orbit interaction [24-27]

## Conclusion

In summary, we have shown that highly mixed Schrodinger cat states can be generated from low purity partially coherent light by unitary evolution in a graded-index waveguide. It is demonstrated that if the mixed cat states generate interference fringes, then in a simple statistical mixture the two wave packets are not coherent and do not interfere. Our analysis also suggests that the cat state is a squeezed state and does not minimize the Heisenberg and generalized Schrodinger-Robertson uncertainty relations. This indicates that the cat state is a nonclassical object. The importance of the effects of long-term decoherence and recoherence of a light beam in the observation of mixed cat states is demonstrated.

## References

[1] Schrodinger, E. Die gegenwдrtige situation in der Quantenmechanik. *Naturwissenschaften* 23, 807-812 (1935).

[2] Monroe, C., Meekhof, D. M., King, B. E. & Wineland, D. J. A "Schrödinger cat" superposition state of an atom. Science 272, 1131-1136 (1996).

[3] Brune, M., Hagley, E., Dreyer, J., Maître, X., Maali, A., Wunderlich, C., Raimond, J.M., Haroche, S. Observing the progressive decoherence of the "meter" in a quantum measurement. Phys. Rev. Lett. 77, 4887-4890 (1996).

[4] Glauber, R.J. Coherent and incoherent states of the radiation field. Phys. Rev. 131, 2766-2788 (1963).

[5] Schrodinger, E. Naturwissenschaften 14, 664-666 (1926).

[6] Dodonov, V. V., Malkin, I.A. and Man'ko, V.I. Even and odd coherent states and excitations of a singular oscillator. Physica 72, 597-615 (1974).

[7] Yurke, B. and Stoler, D. Generating quantum mechanical superpositions of macroscopically distinguishable states via amplitude dispersion. Phys. Rev. Lett. 57, 13-16 (1986).

[8] Deleglise, S., Dotsenko, I., Sayrin, C., Bernu, J., Brune, M., Raimond, J.M. and Haroche, S. Reconstruction of non-classical cavity field states with snapshots of their decoherence. Nature (London) 455, 510-514 (2008).

[9] Ourjoumtsev, A., Jeong, H., Tualle-Brouri, R. and Grangier, P. Generation of optical 'Schrödinger cats' from photon number states. Nature (London) 448, 784-786 (2007).

[10] Petrov, N.I. Macroscopic quantum effects for classical light. Phys. Rev. A 90, 043814 (2014).

[11] Gerry, C. C. & Knight, P. L. *Introductory Quantum Optics* (Cambridge University Press, Cambridge, UK, 2005).

[12] Nielsen, M.A. & Chuang, I. L. *Quantum Computation and Quantum Information* (Cambridge University Press, Cambridge, UK, 2001).

[13] Huyet, G., Franke-Arnold, S., Barnett, S.M. Superposition states at finite temperature. Phys. Rev. A 63, 043812 (2001).

[14] Jeong, H., Ralph, T. C., Transfer of nonclassical properties from a microscopic superposition to macroscopic thermal states in the high temperature limit. Phys. Rev. Lett. 97, 100401 (2006).

[15] Zheng, S.B. Macroscopic superposition and entanglement for displaced thermal fields induced by a single atom. Phys. Rev. A 75, 032114 (2007).

[16] Jeong, H., Ralph, T.C. Quantum superpositions and entanglement of thermal states at high temperatures and their applications to quantum-information processing. Phys. Rev. A 76, 042103 (2007).

[17] Nicacio, F., Maia, R.N.P., Toscano, F., Vallejos, R.O. Phase space structure of generalized Gaussian cat states. Phys. Lett. A 374, 4385-4392 (2010).

[18] Yang, I., Agrenius, T., Usova, V., Romero-Isart, O., Kirchmair, G. Hot Schrödinger cat states. Sci. Adv. 11, eadr4492 (2025).

[19] Mandel, L. & Wolf, E. *Optical Coherence and Quantum Optics* (Cambridge University: Cambridge, MA, USA, 1995).

[20] Krivoshlykov, S.G., Petrov, N.I., Sissakian, I.N. Density-matrix formalism for partially coherent optical fields propagating in slightly inhomogeneous media. Opt. and Quant. Electr. 18, 253-264 (1986).

[21] Starikov, A., Wolf, E.. Coherent-mode representation of Gaussian Schell-model sources and of their radiation fields. J. Opt. Soc. Am. 72, 923-928 (1982).
[22] Petrov, N.I. Long-scale decoherence and recoherence of partially coherent light in a graded-index waveguide. Opt. Lett. 50, 6309-6312 (2025).
[23] Robertson, H. P. The Uncertainty Principle. Phys. Rev. 34, 163-164 (1929).
[24] Petrov, N.I. Spin-orbit and tensor interactions of light in inhomogeneous isotropic media, Phys. Rev. A 88, 023815, 2013.
[25] Petrov, N.I. Splitting levels in a cylindrical dielectric waveguide, Opt. Lett. 38, 2020-2022 (2013).
[26] Petrov, N.I. Vector Laguerre–Gauss beams with polarization-orbital angular momentum entanglement in a graded-index medium, J. Opt. Soc. Am. A 33, 1363-1369 (2016).
[27] Petrov, N.I. Sharp focusing of partially coherent Bessel-correlated beams by a graded-index lens. Opt. Lett. 48, 6048-6051 (2023).